# Raman scattering in superconducting NdO$_{1-x}$F$_x$BiS$_2$ crystals

Yong Tian,[1] Anmin Zhang,[1] Kai Liu,[1] Jianting Ji,[1] Jianzhong Liu,[2] Xiyu Zhu,[2] Hai-Hu Wen,[2] Feng Jin,[1] Xiaoli Ma,[1] Qing-Ming Zhang[1,3*]

[1]Department of Physics, Beijing Key Laboratory of Opto-electronic Functional Materials & Micro-nano Devices, Renmin University of China, Beijing 100872, P.R.China

[2]National Laboratory of Solid State Microstructures and Department of Physics, Innovative Center for Advanced Microstructures, Nanjing University, Nanjing 210093, P.R. China

[3]Department of Physics & astronomy, Innovative Center for Advanced Microstructures, Shanghai Jiao Tong University, Shanghai 200240, P.R. China

## Abstract

The recently discovered layered BiS$_2$-based superconductors have attracted a great deal of interest due to their structural similarity to cuprate and iron-pnictide superconductors. We have performed Raman scattering measurements on two superconducting crystals NdO$_{0.5}$F$_{0.5}$BiS$_2$ (T$_c$ = 4.5 K) and NdO$_{0.7}$F$_{0.3}$BiS$_2$ (T$_c$ = 4.8 K). The observed Raman phonon modes are assigned with the aid of first-principles calculations. The asymmetrical phonon mode around 118 cm$^{-1}$ reveals a small electron-phonon (e-ph) coupling constant λ ~ 0.16, which is insufficient to generate superconductivity at ~ 4.5 K. In the Raman spectra there exists a clear temperature-dependent hump around 100 cm$^{-1}$, which can be well understood in term of inter-band vertical transitions around Fermi surface. The transitions get boosted when the particular rectangular-like Fermi surface meets band splitting caused by spin-orbit coupling. It enables a unique and quantitative insight into the band splitting.

---

* qmzhang@ruc.edu.cn

# Introduction

The discovery of $BiS_2$-based superconductors has inspired a new wave of research[1-5]. Analogous to cuprate and iron-pnictide superconductors, the new superconductors have two-dimensional conducting $BiS_2$ layers as key building blocks, which are considered to be responsible for superconductivity (SC). Actually the $BiS_2$-based compounds are very similar to FeSe superconductors in structure and exhibit semiconducting behavior at low temperatures[6, 7]. As the most common elementary excitations in solids, lattices vibrations are vital to study some fundamental physical properties and interactions, such as mechanical and thermal properties, electron-phonon coupling and other phonon-involved scattering processes. Phonon dispersion in the materials has been initially studied by neutron scattering[8]. And Raman measurements have also been made recently[9]. However, a comprehensive assignment of the observed Raman modes is still lacking. Furthermore, many key electronic properties, such as e-ph coupling and low-energy excitations, are waiting to be unraveled by Raman scattering and other optical methods.

The electronic properties of $REO_{1-x}F_xBiS_2$ (RE = rare earth) compounds have received many theoretical and experimental investigations. First-principles calculations predicted a Fermi surface showing a nesting along ($\pi$, $\pi$, 0) direction at half filling. This may induce phonon instability or charge density wave (CDW), which implies that superconductivity in the $BiS_2$ system may be driven by strong e-ph coupling[10-12]. Actually no sign of CDW transition was observed by infrared optical spectroscopy[13], and neutron scattering measurements show no significant change in phonon density cross the SC transition[8]. Rectangular-like Fermi pockets around X points have been revealed by angle resolved photoemission spectroscopy (ARPES) experiments. The measured band filling is smaller than the calculated one, against the picture of Fermi surface nesting[14, 15]. The pairing mechanism in the $BiS_2$-based superconductors is still under debate and requires further theoretical and experimental studies.

Due to the particular rectangular-like Fermi surface, the delicate electronic structure around Fermi surface becomes crucial to the understanding of electronic properties of $BiS_2$-based compounds. A multi-band or shallow band-edge effect, observed by Hall measurements, seems to be related to SC pairing in the $CeO_{1-x}F_xBiS_2$[16]. And ARPES has observed a band splitting due to spin-orbit coupling (SOC) in $NdO_{0.5}F_{0.5}BiS_2$[14]. Probing low-energy excitations around Fermi surface is still a one of the key issues in $BiS_2$-based superconductors and is highly desired.

In this work, we have performed Raman scattering experiments on SC single crystals $NdO_{0.5}F_{0.5}BiS_2$ ($T_c$=4.5K) and $NdO_{0.7}F_{0.3}BiS_2$ ($T_c$=4.8K). The observed Raman modes are assigned with the assistance of first-principles calculations. The $A_{1g}$ mode at 118 cm$^{-1}$ exhibits an asymmetrical lineshape which is attributed to Breit-Wigner-Fano (BWF) resonance (or Fano resonance) driven by e-ph coupling[17, 18]. The fitting of Fano lineshape gives the strength of e-ph coupling constant $\lambda \sim 0.16$, which gives a near-zero $T_c$ using the density of states (DOS) at Fermi surface N ($E_F$) derived from electronic specific heat coefficient[19]. This may suggest that the paring glue in the $BiS_2$ system is beyond e-ph coupling. Furthermore, the rectangular-like Fermi surface plus band splitting due to SOC, enhances the inter-band vertical transitions around Fermi surface. The transitions give rise to a clear temperature-dependent electronic hump in cross-polarization Raman channel, which allows a quantitative study on band splitting.

**Experiments and methods**

Superconducting single crystals $NdO_{0.5}F_{0.5}BiS_2$ ($T_c$ = 4.5 K) and $NdO_{0.7}F_{0.3}BiS_2$ ($T_c$ = 4.8 K) were grown using KCl flux and the details of crystal growth and characterization can be found elsewhere[20, 21]. The sample size is typically about 0.5×0.5×0.1 mm. The crystals can be easily cleaved to obtain flat and shining *ab* planes. To exclude possible impurity effect on surface, we have repeated the cleavage for the crystals at least two times before they were placed into the cryostat. For each cleavage, Raman spectra were recorded and compared. And no sign of impurity is seen in the spectra (Supplementary materials). It is in agreement with the reported structural characterizations[7]. Raman measurements were performed in a high-vacuum closed-cycle cryostat with anti-vibration design (10~300 K, ~$10^{-8}$ mbar). Raman spectra were collected with a HR800 spectrometer (Jobin Yvon) equipped with liquid nitrogen cooled CCD and volume Bragg gratings and micro-Raman backscattering configuration was adopted. A 632.8 nm laser was used with a spot of ~5 microns focused on sample surface, and the laser power less than 1 mW was sustained during measurements to avoid overheating.

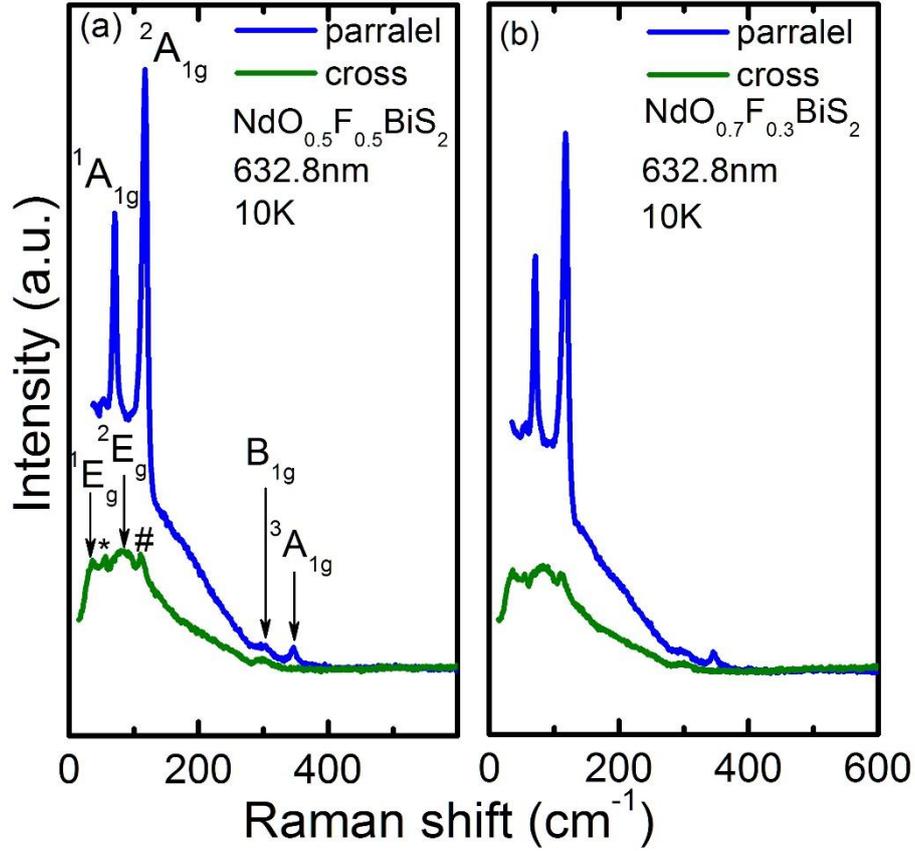

Fig. 1. Raman spectra of $NdO_{1-x}F_xBiS_2$ (x = 0.5, 0.3) in cross and parallel channels at 10K. The assigned Raman modes are indicated with arrows. The small peaks marked with ∗ and # may originate from the leakage of infrared mode and $^2A_{1g}$ mode, respectively.

In order to identify the Raman active modes found in experiments, we had carried out first-principles calculations by using the QUANTUM-ESPRESSO (QE) package[22]. The ultra-soft pseudopotentials[23] were employed to describe the electron-ion interactions. The generalized gradient approximation (GGA) of Perdew-Burke-Ernzerhof (PBE) [24] formalism was adopted for the exchange and correlation functional. The kinetic energy cutoff and the charge density cutoff for the plane-wave basis were 800 eV and 8000 eV, respectively. For the Brillouin-zone integration, the *k* points were sampled on a 12X12X4 grid. The Fermi surface was broadened using the gaussian smearing technique with a width of 0.05 eV. Variable-cell calculations were performed until zero pressure was reached and all of the forces acting on atoms were less than 0.01 eV/Å.

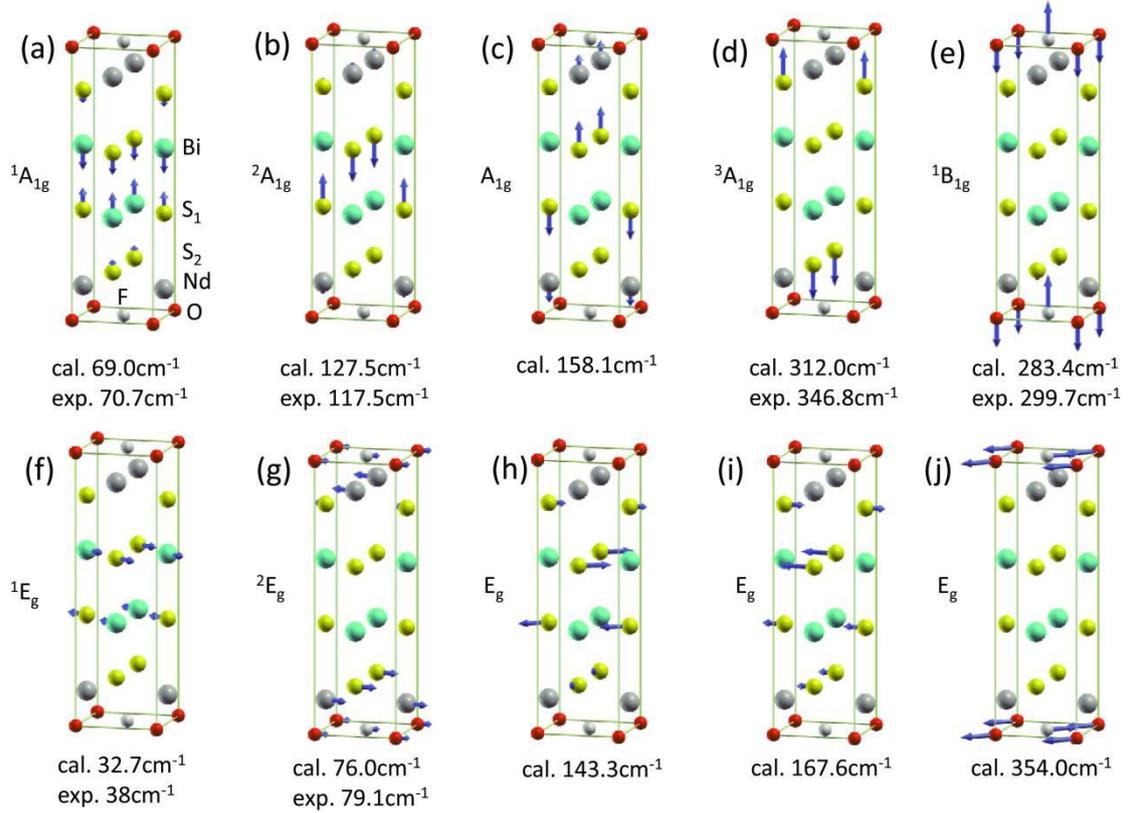

Fig. 2. Vibration patterns of Raman-active modes in NdO$_{0.5}$F$_{0.5}$BiS$_2$. The calculated and experimental mode frequencies are listed below each pattern.

## Results and discussions

### I. Assignments of Raman modes

Low-temperature Raman spectra of NdO$_{0.5}$F$_{0.5}$BiS$_2$ and NdO$_{0.7}$F$_{0.3}$BiS$_2$ are shown in Fig. 1. The spectra exhibit a high similarity between both crystals in peak position and shape. This implies that the actual composition of the two samples may be very close though their nominal fluorine contents seem quite different. It is in accord with the fact that both crystals have very close SC transition temperatures. Symmetry analysis shows that there are ten Raman-active modes ($4A_{1g}+B_{1g}+5E_g$) at gamma point. The corresponding vibration patterns are illustrated in Fig. 2. The observed Raman modes are well assigned in combination with first-principles calculations as shown in Fig. 2. And the calculated phonon energies are close to those from Ref. 8 and summarized in Table I.

Among the observed Raman modes, the two peaks located at 70 cm$^{-1}$ and 117.5 cm$^{-1}$, only visible in parallel polarization channel, are identified as $A_{1g}$ symmetry. The $^2A_{1g}$ phonon mode exhibits a clear asymmetric lineshape, indicating an obvious e-ph coupling. A broad low-frequency hump in cross polarization channel is observed in both samples. We leave detailed discussions on them in the following sections. The small peaks around 38 cm$^{-1}$ and 79 cm$^{-1}$ in cross channel, is attributed to $E_g$ modes according to first-principles calculations (see Table. 1). The modes are normally prohibited in the present scattering geometry but may leak into the measurements on *ab* planes, perhaps due to a slight tilt of sample surface. There are two unidentified tiny peaks in cross channel. One is the peak around 111.5 cm$^{-1}$ (marked by #) which can be reasonably attributed to intensity leakage of the strongest $^2A_{1g}$ mode in parallel channel. And the other peak at 56 cm$^{-1}$ (marked by *), may be a leakage of Raman-inactive mode. Actually around the energy, a peak of phonon density of states has been observed by neutron scattering measurements.[8] The $^2E_g$ and $B_{1g}$ modes come from the cooperative vibrations of O/F atoms. Both modes seem clearly boarder than others. This is consistent with O/F disorder.

Table I Assignments of the observed Raman modes. The calculated phonon frequencies of NdO$_{0.5}$F$_{0.5}$BiS$_2$ are in good agreement with experiments. S$_1$ and S$_2$ denote sulfur atoms in BiS plane and out of plane, respectively. (See Fig. 2)

| Symmetry | Calc. Freq. | Exp. Freq. | Atoms |
|---|---|---|---|
| $^1E_g$ | 32.7 | 38 | Bi and S$_1$ |
| $^1A_{1g}$ | 69.0 | 70.7 | Bi, S$_1$ and S$_2$ |
| $^2E_g$ | 76.0 | 79.1 | S$_2$, Nd and O/F |
| $^2A_{1g}$ | 127.5 | 117.5 | Nd and S$_1$ |
| $E_g$ | 143.3 | | S$_1$ and S$_2$ |
| $A_{1g}$ | 158.1 | | Nd and S$_1$ |
| $E_g$ | 167.6 | | S$_1$ and S$_2$ |
| $^1B_{1g}$ | 283.4 | 299.7 | O/F |
| $^3A_{1g}$ | 312.0 | 346.8 | S$_2$ |
| $E_g$ | 354.0 | | O/F |

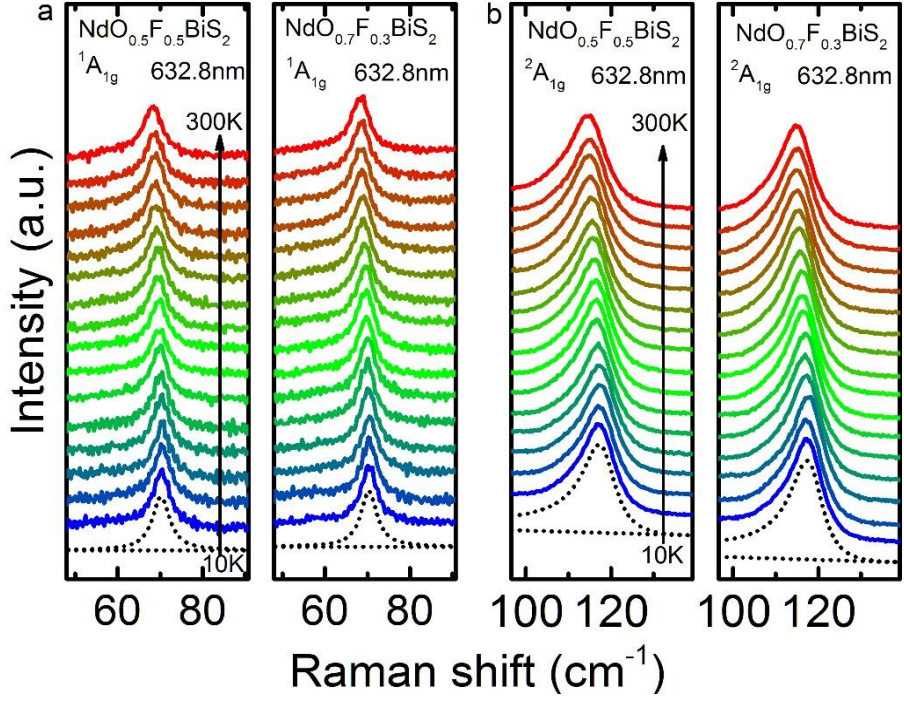

Fig. 3. Temperature evolution of $^1A_{1g}$ (a) and $^2A_{1g}$ (b) modes in $NdO_{0.5}F_{0.5}BiS_2$ and $NdO_{0.7}F_{0.3}BiS_2$. The dashed peaks at the bottoms are Lorentzian fitting (a) and Fano fitting (b).

## II. Electron-phonon coupling

The temperature dependence of the two strongest phonon peaks, $^1A_{1g}$ and $^2A_{1g}$, is depicted in Fig. 3. Both modes indicate a smooth temperature evolution and no observable anomalies are seen in the temperature range of 10 to 300 K. However, their line shapes are clearly different. $^1A_{1g}$ mode is a well-defined Lorentzian-type peak, while on the other hand, $^2A_{1g}$ displays an asymmetric Fano-like shape (see below). The fitting curves using symmetric and asymmetric functions for the two modes, are illustrated at the bottoms of Fig. 3a&3b. Fig. 4 further shows the temperature dependence of phonon frequencies. Both modes can be well described by anharmonic phonon process

$$\omega(T) = \omega_0 - C \times \left(1 + \frac{2}{e^{\frac{\hbar\omega_0}{2k_BT}} - 1}\right) \qquad (1)$$

where $\omega_0$, $k_B$ and $C$ are the phonon frequency extrapolated to 0 K, Boltzmann constant and fitting constant, respectively. The smooth temperature evolution suggests the absence of structure phase change over the measured temperature range, which is in agreement with the experimental facts of no CDW transition[13]. The frequencies of the two modes extrapolated to 0 K (70.5 cm$^{-1}$ and 118.94 cm$^{-1}$ respectively) are almost the same in both crystals, which may imply a close composition in nominal $NdO_{0.5}F_{0.5}BiS_2$ and $NdO_{0.7}F_{0.3}BiS_2$ as mentioned above.

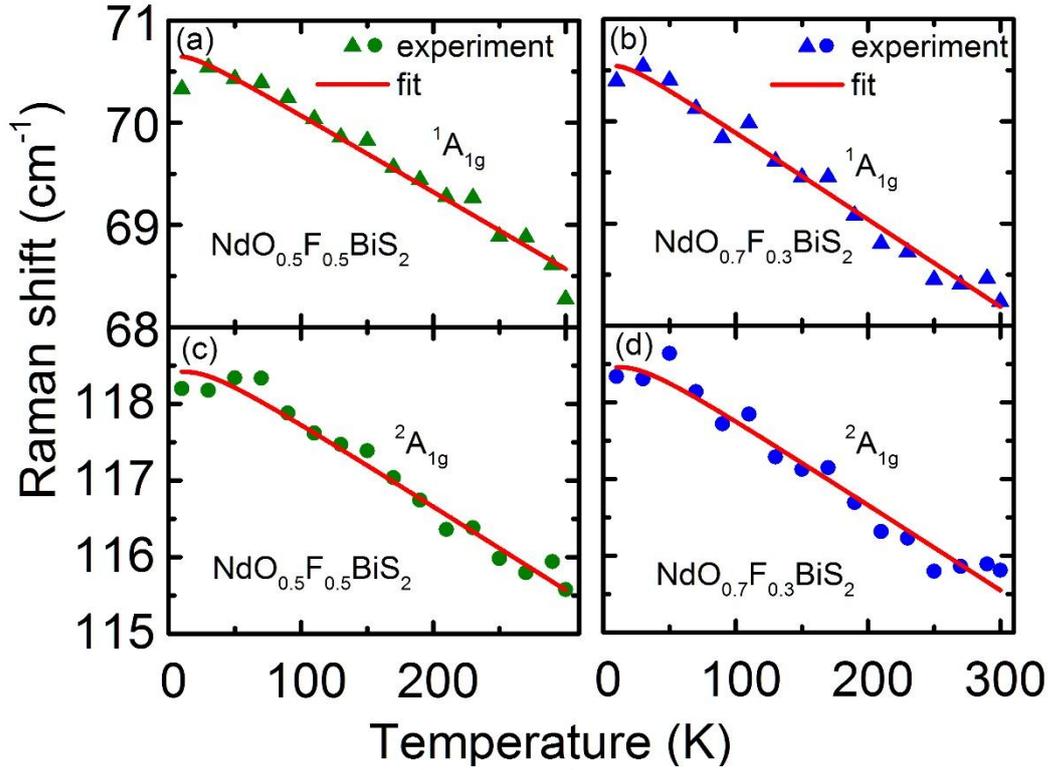

Fig. 4. Temperature dependence of $^1A_{1g}$ (a) and $^2A_{1g}$ mode frequencies in $NdO_{0.5}F_{0.5}BiS_2$ (a&c) and $NdO_{0.7}F_{0.3}BiS_2$ (b&d). The red curves are anharmonic fittings using formula (1).

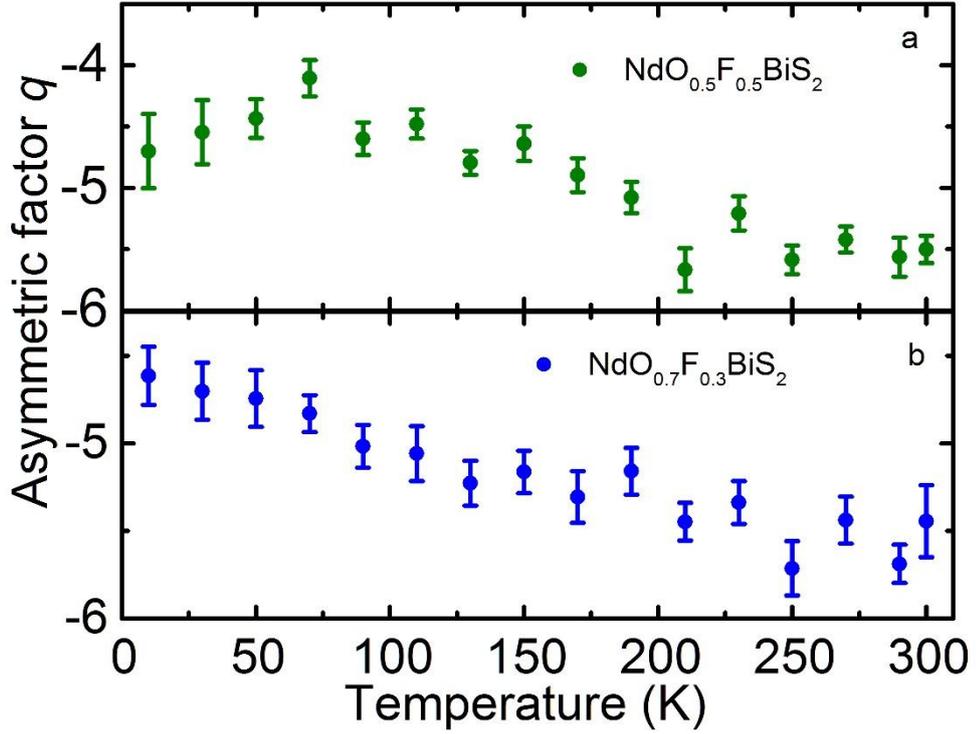

Fig. 5. Temperature dependence of asymmetric factor $q$ of $^2A_{1g}$ mode for $NdO_{0.5}F_{0.5}BiS_2$ (a) and $NdO_{0.7}F_{0.3}BiS_2$ (b).

The asymmetric $^2A_{1g}$ exhibits a typical Breit-Wigner-Fano (BWF) behavior, which indicates the interaction between discrete and continuous excitations. It is e-ph coupling in our case. Fano equation reads[17,28]

$$I(\varepsilon, q) = \frac{I_0 \times (q+\varepsilon)^2}{1+\varepsilon^2} \qquad (2)$$
$$\varepsilon = (\omega - \omega_0)/\Gamma$$

where $I_0$ and $\Gamma$ are scaling factor and full width at half maximum (FWHM), $q$ is an inverse measure of asymmetry, which means that a larger q actually corresponds to a more symmetric peak. A peak will return to a pure Lorentzian shape at the limit of infinite $q$. The temperature dependence of extracted q is shown in Fig. 5. The value of q lies in the range of -4 to -6 and has a slight increase with decreasing temperatures. This means that a finite e-ph coupling always exists above SC transition temperatures.

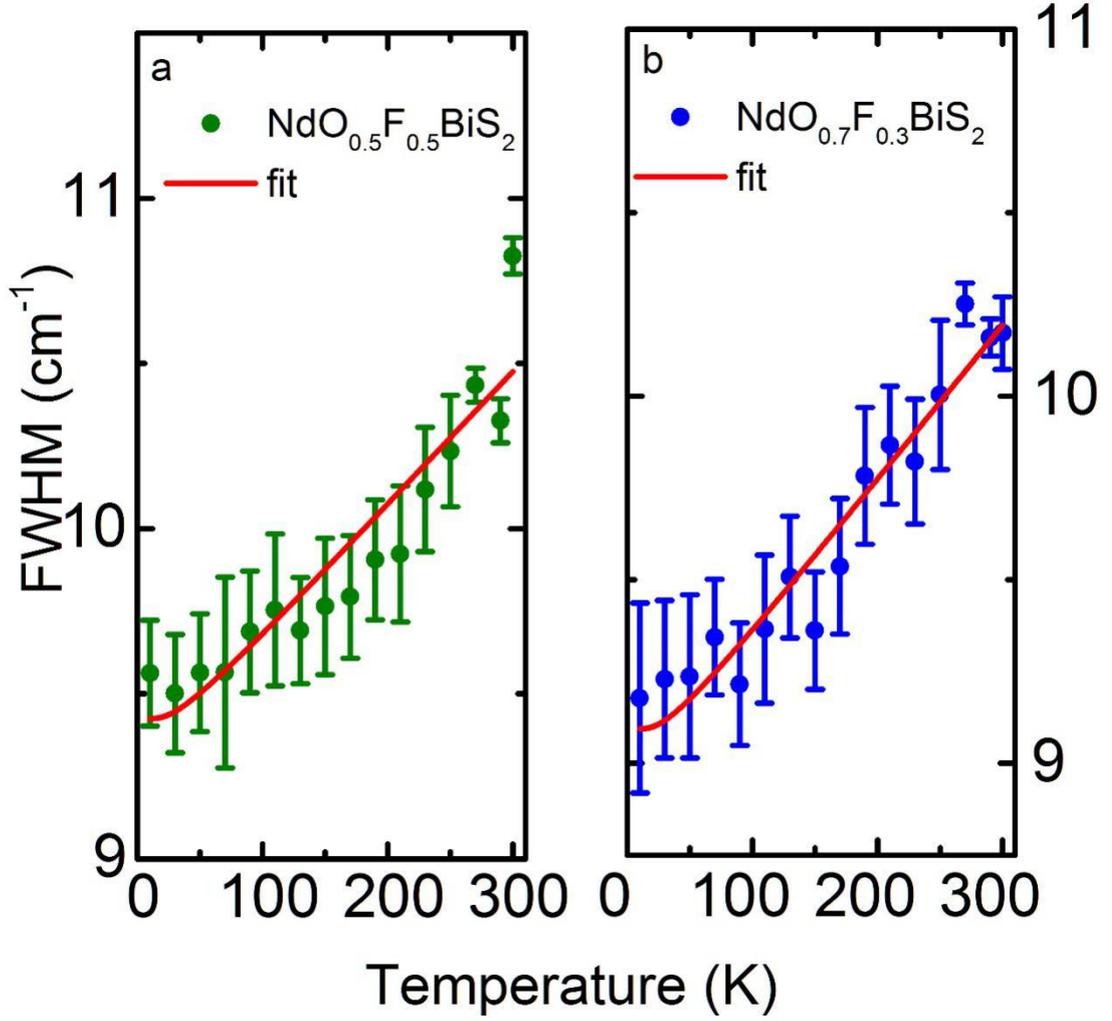

Fig. 6. Temperature dependence of $^2A_{1g}$ phonon widths in $NdO_{0.5}F_{0.5}BiS_2$ (a) and $NdO_{0.7}F_{0.3}BiS_2$ (b). The red curves are anharmonic fittings using formula (2).

The e-ph coupling constant $\lambda$ can be determined by phonon broadening $\Gamma^{e\text{-}ph}$ [29]. The broadening is obtained from the measured FWHM, the jump of phonon frequencies at SC transition[29, 30], or the asymmetry factor $q$[31]. Actually it is quite complicated to derive $\Gamma^{e\text{-}ph}$ from $q$. Measuring the frequency jump cross SC transition seems a relatively reliable way to estimate intrinsic phonon broadening. Unfortunately such jump is not available in our case. We will use $\Gamma^{e\text{-}ph}$ extracted from the measured linewidths. Basically the total linewidth $\Gamma$ is contributed by anharmonic phonon decay, temperature-independent impurity scattering and e-ph coupling. Accordingly, the linewidth can be described as[31],

$$\Gamma = \Gamma_{anh}^0\left(1+\frac{2}{\exp(\hbar\omega_p/2k_BT)-1}\right)+\Gamma_b \quad (3)$$

where $\Gamma_{anh}^0$ is the anharmonic phonon broadening extrapolated to 0 K and $\Gamma_b$ is the broadening caused by e-ph coupling and impurity scattering. Approximately we can take the width of symmetric $^1A_{1g}$ mode extrapolated to 0 K as the contribution from impurity scattering, i.e, $\Gamma^{e-p} \approx \Gamma^b - \Gamma(^1A_{1g},0K) \approx 4.0 cm^{-1}$ for $NdO_{0.5}F_{0.5}BiS_2$ (Fig. 6(a)). This allows to estimate an intrinsic broadening from e-ph coupling $\Gamma^{e-ph} \sim 5.3$ cm$^{-1}$, The experimental density of states (DOS) at Fermi surface, $N_{exp}(E_f) \approx 5.5$ states/eV/spin/formula, can be derived from electronic specific heat measurements[19]. The e-ph coupling constant is expressed as[29]

$$\lambda = \Gamma^{e-ph}\Big/\pi N(E_F)\hbar\omega_0^2 \quad (4)$$

where $N(E_f)$ is the bare density of state at Fermi energy and given by $N_{exp}(E_f)/(1+\lambda)$[32, 33]. Eq. (4) gives $\lambda \sim 0.16$ with $\omega_0 = 118.9$ cm$^{-1}$. The derived e-ph coupling constant is supported by first-principles calculations, which give a close value for the same mode.[8] The coupling strength leads to a $T_c \sim 0$ K using Allen-Dynes formula[34, 35]. Even the original linewidth ~9.3 cm$^{-1}$ without subtracting impurity broadening,, only produces a $\lambda \sim 0.47$ and $T_c \sim 1.1$ K. The value is still smaller than the experimental ones, which implies that the pairing glue for the new superconductors may not be the e-ph coupling. We note that in ref. 9 the authors employed a calculated DOS rather than an experimental one, which outputs an e-ph coupling constant of ~ 0.68 and a $T_c$ of ~ 5 K.

## III. Inter-band transitions

Now we turn to the discussions on the low-frequency hump mentioned above. Fig. 7 shows Raman response (imaginary part of Raman susceptibility) in cross polarization channel. A clear hump appears around 80 cm$^{-1}$ at 10 K and smoothly shifts to 166 cm$^{-1}$ at 300 K. Actually the hump can be also clearly seen in parallel channel but superimposed by strong phonon modes (Fig. 1) . A phononic origin for the hump can be excluded because of its non-phononic-like shape, very large width and the unusual softening with decreasing temperatures. For the same reasons, the possibility that it comes from phonon signals of impurities, can be also ruled out. And structural characterizations demonstrate that the impurities, if existing, must have a negligible concentration in the crystals[7]. ARPES and optical measurements indicate that the plasma frequency in this system is ~2.1 eV, far beyond the observed characteristic energy ~ 0.15 eV[13, 14]. Magnetic excitations also seem unlikely to be the origin of the hump. Rare-earth $Nd^{3+}$ are the only magnetic ions in the compounds. There is no observation of spin ordering of $Nd^{3+}$ ions so far. Even if $Nd^{3+}$ spins are ordered, generally it would occur at very low temperatures (~10 K). Thus, the above discussions lead to the possibility that the hump is contributed by electronic excitations at Fermi surface.

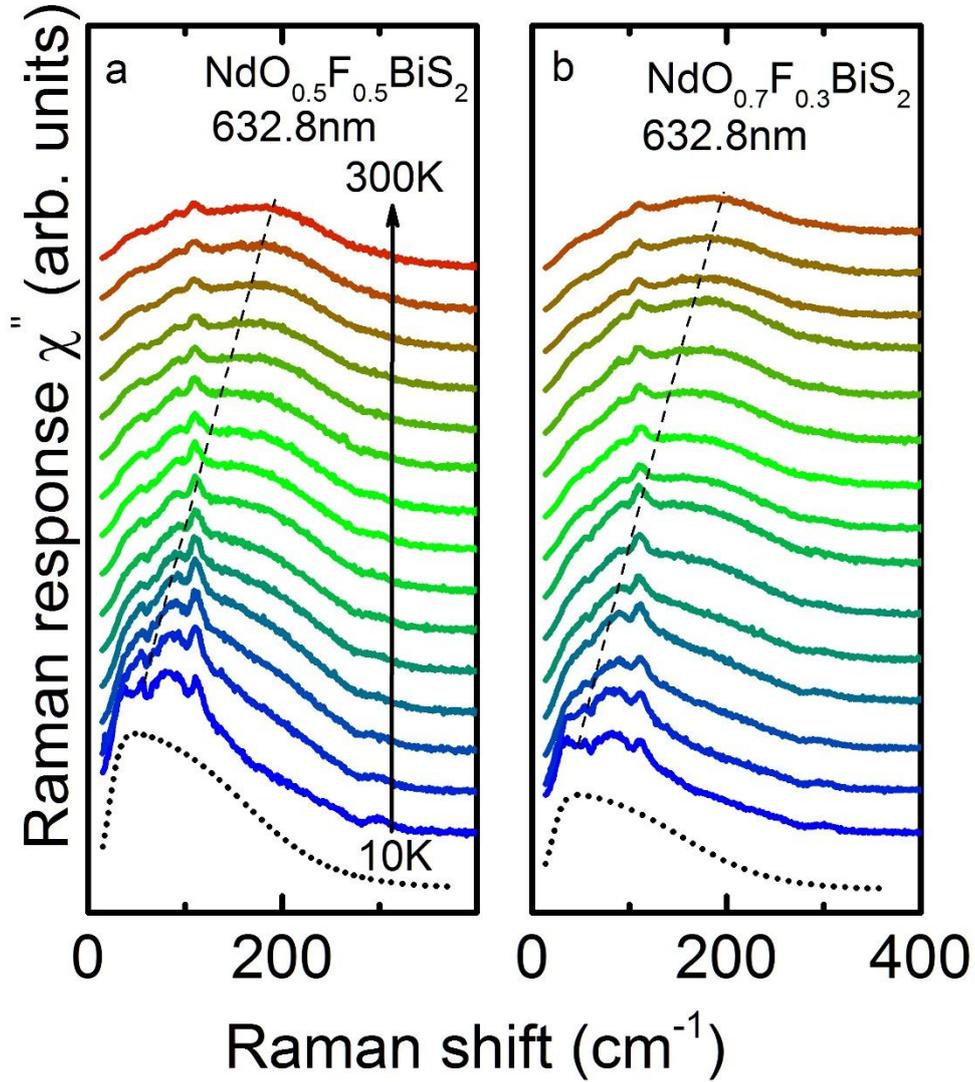

Fig. 7. Temperature evolution of low-energy hump in cross channel in $NdO_{0.5}F_{0.5}BiS_2$ (a) and $NdO_{0.7}F_{0.3}BiS_2$ (b). The dashed lines at the bottoms are fitting curves using (5).

Generally electronic Raman scattering (ERS) requires a finite momentum transfer, which is usually provided by phonons or impurities due to near-zero momentum of incident photons. But a special case is seen in the new $BiS_2$-based superconductors. ARPES measurements reveal a finite band splitting caused by SOC at Fermi level and two separate rectangular-like Fermi surfaces around X points[14, 15]. The band splitting allows a vertical inter-band hopping around Fermi level without the assistance of phonons or impurities, as illustrated in Fig. 8. The same vertical hopping occurs not just at one k point, but along the longer side (Γ-X) of rectangular Fermi surfaces. This dramatically boosts ERS signals, which actually contribute the observed low-frequency hump. A further quantitative analysis can be made based on the picture.

For the vertical electron hopping between the two split bands, Raman response $\chi''(\omega)$ is in general proportional to the hopping probability. Thus Raman response can be written as

$$\chi''(\omega) \sim N_{E_F}^2 \times f(E^{lower}) \times [1 - f(E^{upper})]$$

$$= N_{E_F}^2 \times \frac{1}{\exp\left[(E^{lower} - E_F)/k_B T\right] + 1} \times \left[1 - \frac{1}{\exp\left[(E^{upper} - E_F)/k_B T\right] + 1}\right] \quad (5)$$

where $E^{lower}$ and $E^{upper}$ are the energies of electrons in lower and upper bands, respectively. Both energies can be linearly expanded around Fermi level, i.e., $E^{lower} \approx \hbar \cdot v_F^{lower} \cdot \Delta k^{lower} + C_1$ and $E^{upper} \approx \hbar \cdot v_F^{upper} \cdot \Delta k^{upper} + C_2$ where $\Delta k^{lower}/\Delta k^{upper}$ is electron momentum relative to Fermi momentum $k_F^{lower}/k_F^{upper}$. Now Raman shift $\omega$ corresponds to the transfer energy related to the vertical transition, i.e., $\hbar\omega = E^{upper} - E^{lower}$. Taking the parameters into (5), we can make a good fitting for

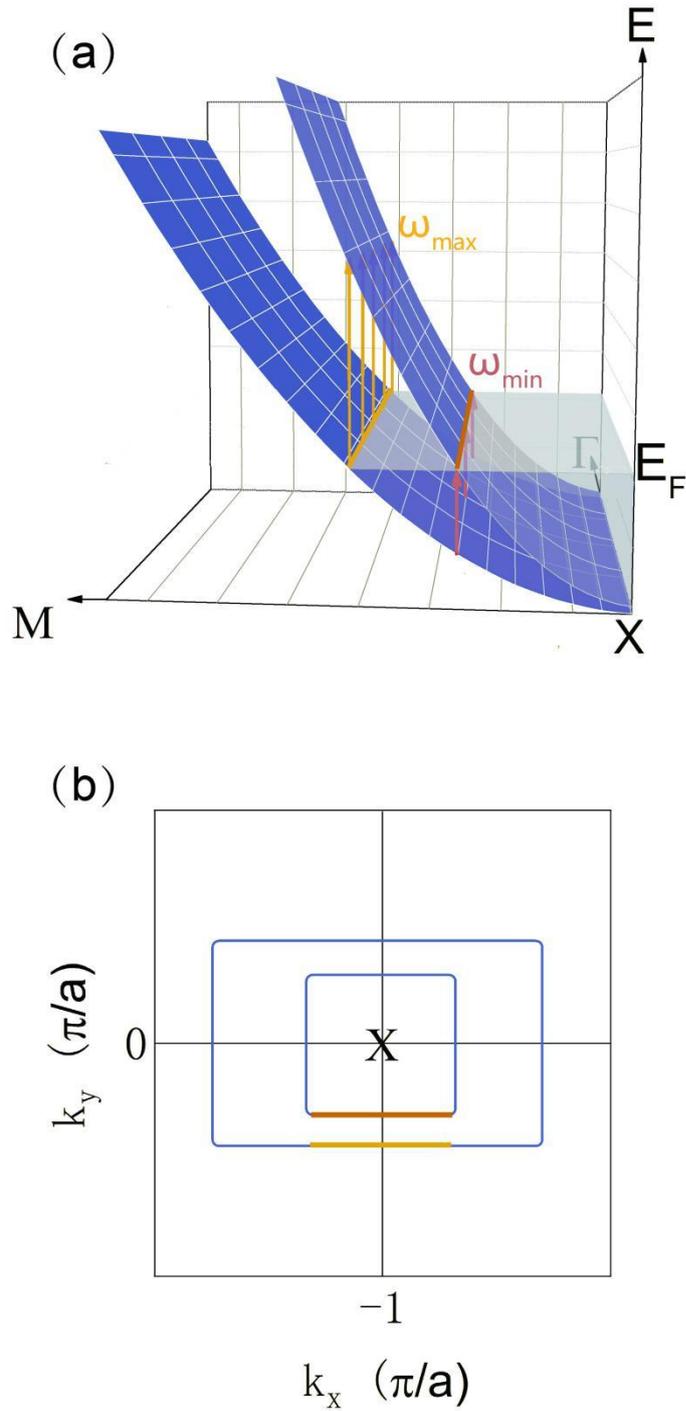

Fig. 8. (a) schematic of band structure near Fermi level in $NdO_{1-x}F_xBiS_2$ (x = 0.5, 0.3). The two bands shown here, come from a splitting by spin-orbit coupling. The vertical hopping, visible in Raman channel, is illustrated by yellow and red arrows. $\omega_{max}$ and $\omega_{min}$ are the maximum and minimum vertical hopping energies at zero temperature. (b) Top view of schematic Fermi surface along $k_z$. The two colored lines correspond to the Fermi surface segments in (a) marked with the same color lines.

. Taking the parameters into (5), we can make a good fitting for the hump, as shown at the bottom of Fig. 7a & 7b. The fitting allows us to define two characteristic energy scales, $\omega_{min}$ and $\omega_{max}$, at which the hump takes half of its maximum. As depicted in Fig. 8, the energies correspond to two particular vertical hopping processes in which initial/final states are exactly located at Fermi level. They can be regarded as a good measure of band splitting. At 10 K, the fitting gives $\omega_{min}$ ~ 24.79cm$^{-1}$ (~3 meV) for NdO$_{0.5}$F$_{0.5}$BiS$_2$ and 23.62cm$^{-1}$ (~3 meV) for NdO$_{0.7}$F$_{0.3}$BiS$_2$; $\omega_{max}$ ~ 169.3cm$^{-1}$ (~21.0 meV) for NdO$_{0.5}$F$_{0.5}$BiS$_2$ and 158.2cm$^{-1}$ (~19.6 meV) for NdO$_{0.7}$F$_{0.3}$BiS$_2$. The values are in accord with the results given by ARPES[14, 15], which demonstrates that the band splitting around X points lies in the range of 5 ~ 30 meV.

## Summary

In summary, we have measured polarized Raman scattering spectra of BiS-based SC single crystal NdO$_{0.5}$F$_{0.5}$BiS$_2$ (T$_c$ = 4.5K) and NdO$_{0.7}$F$_{0.3}$BiS$_2$ (T$_c$ = 4.8K) from 10K to 300K. The observed Raman modes are assigned with the aid of first-principles calculations. And an investigation on the Fano-type A$_{1g}$ mode gives an electron-phonon coupling constant λ ~ 0.16, which is not strong enough to produce superconductivity at ~4.5 K. We further observe a low-energy electronic hump. It can be quantitatively explained by the vertical transitions around Fermi surface between the two bands split by spin-orbit coupling. The transitions are dramatically enhanced due to the particular rectangular-like Fermi surface, and hence provides a unique window to quantitatively probe the band splitting.

## Acknowledgements

This work was supported by the Ministry of Science and Technology of China (973 projects: 2011CBA00112，2011CBA00102, and 2012CB921701) and the NSF of China (Grant No.: 11174367 & 11474357). Q.M.Z. was supported by the Fundamental Research Funds for the Central Universities, and the Research Funds of Renmin University of China.

Supplementary information

To exclude the impurity and surface effect, we have made various Raman measurements on different crystals which are repeatedly cleaved. The raw Raman spectra of four crystals are shown in Fig. S1. Each crystal is cleaved at least twice and Raman spectrum was recorded immediately after each cleavage. All the spectra show almost the same features and no substantial difference is seen among them. This indicates that impurity and/or surface effect is negligible in our case. The shoulder below 210 cm$^{-1}$ comes from electron hopping between two split bands near Fermi surface (see Section III in context).

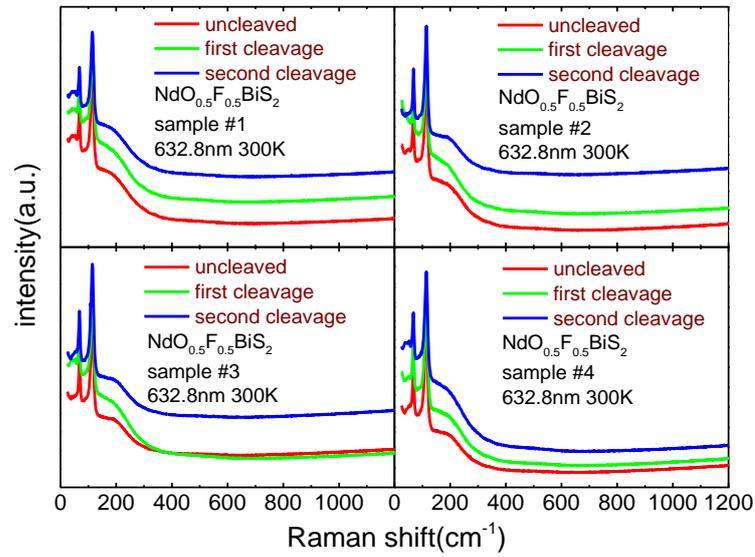

FIG. S1: Raman spectra of four NdO$_{0.5}$F$_{0.5}$BiS$_2$ crystals with multiple cleavages.